\documentclass{radioeng}
\usepackage[czech,english]{babel}

\usepackage[cp1250]{inputenc}
\usepackage{graphicx} % To include and scale graphics
\usepackage{color,soul}
\usepackage{cite}
\usepackage{amsmath,bm}
\usepackage{amssymb}
\usepackage{tabularx}
\usepackage{textcomp}

\usepackage{listings}  %To include listings

\newcommand{\area}{
CIRCUITS
} % CATEGORY

\begin{document}

\setcounter{firstpage}{1}

\rengHeader{}{}{
... %1
%JUNE %2
%SEPTEMBER %3
%DECEMBER %4
}{V. J. DOWLING, V. A. SLIPKO, Y. V. PERSHIN, MODELING NETWORKS OF PROBABILISTIC MEMRISTORS IN SPICE \dots}

\rengTitle{Modeling networks of probabilistic memristors in SPICE}

\rengNames{Vincent~J.~DOWLING $^\mathit{1}$, Valeriy~A.~SLIPKO $^\mathit{2}$, Yuriy~V.~PERSHIN $^\mathit{1}$}

\rengAffil{$^1$ Department of Physics and Astronomy, University of South Carolina, Columbia, SC 29208 USA \\
$^2$ Institute of Physics, Opole University, Opole 45-052, Poland}

\rengMail{pershin@physics.sc.edu}

\rengReceived{...}{...} % submitted accepted (do not fill out)

%%%%%%%%%%%%%%%%%%%%%%%%%%%%%%%%%%%%%%%%%%%%%%%%%%%%%%%%%%%%%%%
% Abstract, keywords
%%%%%%%%%%%%%%%%%%%%%%%%%%%%%%%%%%%%%%%%%%%%%%%%%%%%%%%%%%%%%%%
\begin{multicols}{2}

\begin{rengAbstract}
Efficient simulation of probabilistic memristors and their networks requires novel modeling approaches. One major departure from the conventional memristor modeling is based on a master equation for the occupation probabilities of network states [arXiv:2003.11011 (2020)]. In the present article, we show how to implement such master equations in SPICE -- a general-purpose circuit simulation program. In the case studies, we simulate
the dynamics of ac-driven probabilistic binary and multi-state memristors, and dc-driven networks of probabilistic binary and  multi-state memristors. Our SPICE results are in perfect agreement with known analytical solutions. Examples of LTspice codes are included.
\end{rengAbstract}

\rengKeywords{Memristors, SPICE, networks, probabilistic computing}

\rengSection{Introduction}

SPICE simulation~\cite{vladimirescu1994spice,kundert2006designer} is a powerful tool in the hands of an electrical engineer. In the last decade significant progress has been made in developing
SPICE models of memristive devices~\cite{Biolek2009-1,Benderli09a,rak10a,sharifi10a,Zhang10a,pershin13c,Biolek13a,da2014two,vourkas2015spice,li2015memristor,Biolek16a,garcia2016spice}, as well as memcapacitive and meminductive elements~\cite{Biolek13a,Biolek2009-2}. The common feature of these previous approaches is the use of differential equations to describe the deterministic evolution of internal state(s) of memory devices~\cite{chua76a,diventra09a}.

However, there is a strong indication that the deterministic description fails when applied at least to certain realizations of resistors with memory~\cite{jo2009programmable,gaba2013stochastic,gaba2014memristive}. In particular, it was shown experimentally that when a constant voltage is applied to such devices, their state changes in a step-like fashion at random times. In one group of devices, a Poisson distribution of switching times was observed~\cite{jo2009programmable,gaba2013stochastic,gaba2014memristive}. Furthermore, another group of devices is characterized by a log-normal distribution~\cite{Medeiros_Ribeiro_2011}. Several theoretical models were pushed forward to account for the randmoness in the memristor
switching~\cite{menzel2014statistical,Naous16a}.

The dynamics of networks with discrete-state memristors can be imagined as a sequence of transitions between network states. Recently,
we have introduced a master equation approach for the occupation probabilities of the network states~\cite{dowling2020probabilistic}
that can be used to describe circuits that include binary and multi-state memristors, resistors, voltage and current sources, and possibly some other components~\footnote{A generalized approach is needed for circuits combining probabilistic memristors and capacitors/inductors.}.
In this previous work~\cite{dowling2020probabilistic}, the solution of the master equation was found analytically for networks of $N$ in-series/in-parallel connected binary memristors driven by a constant voltage source. It has been demonstrated in Ref.~\cite{dowling2020probabilistic} that the master equation solution allows to calculate many quantities of interest including various mean switching times, mean current, resistance, etc.

There are two major advantages of the master equation compared to stochastic/Monte Carlo simulations: $i$) in principle, the master equation can be solved analytically (see Ref.~\cite{dowling2020probabilistic} for examples), and $ii$) using the master equation, many network characteristics can be found in a single calculation without the need for averaging. In the case of symmetries in the circuit, the additional benefit of the master equation is its compactness. This means that a single degree of freedom is required to describe equivalent circuit configurations.

In this article (which is our second work in a series dedicated to probabilistic memristive networks), we introduce a methodology to simulate the probabilistic memristive networks in SPICE. The paper is organized as follows. We start with an overview of the master equation in relation to binary and multi-state probabilistic memristor networks. This is followed by a description of the SPICE implementation scheme supplemented by several examples. In particular, we consider individual probabilistic binary and tri-state memristors driven by ac-voltage, and dc-driven networks thereof. LTspice codes for some of our examples are provided in the Appendix.

The approach presented in this work is relatively general and can be used to model networks combining resistors, probabilistic memristors, constant and time-dependent voltage and current sources. The application of the master equation to probabilistic memristor networks is a paradigm change in the probabilistic memristor modeling, and its SPICE implementation makes it affordable to students and researchers working in the field.

\rengSection{Probabilistic memristors and master equation}

\rengSubsection{Binary memristors}

Binary memristors are characterized by two resistance states, $R_{on}$ and $R_{off}$ (with $R_{on}<R_{off}$) corresponding to the states 1 (on) and 0 (off). The switching between these states is defined by a probabilistic law with voltage-dependent switching rates (inverses of the mean switching times) given by~\cite{jo2009programmable,gaba2013stochastic,gaba2014memristive}
 \begin{eqnarray}
 \gamma_{0\rightarrow 1}(V)=\left\{ \begin{array}{cl}
\left( \tau_{01} e^{-V/V_{01}}\right)^{-1},& V>0 \\
0 & \textnormal{otherwise}
\end{array}\right. \; , \label{eq:gamma01}\\
 \gamma_{1\rightarrow 0}(V)=\left\{ \begin{array}{cl}
\left( \tau_{10} e^{-|V|/V_{10}}\right)^{-1},& V<0 \\
0 & \textnormal{otherwise}
\end{array}\right. \; . \label{eq:gamma10}
 \end{eqnarray}

\noindent Here, $\tau_{01(10)}$ and $V_{01(10)}$ are constants and $V$ is the voltage across the device.
For a memristor in state 0, the probability to switch to state 1 within small time interval $\Delta t$ is  $\gamma_{0\rightarrow 1}(V) \Delta t$. The probability of swithching from 1 to 0 is defined similarly.

The master equation is written with regard to the occupation probabilities of network states. The network state is defined by a specific combination of the off- and on-states of memristors. For a system containing $N$ binary memristors, there exists $2^N$ such states. The network evolution consists of a chain of consecutive switchings of memristors (simultaneous switchings can be neglected). On average, such a process is described by the master equation with form
\begin{equation}
\frac{\textnormal{d}p_{\Theta}(t)}{\textnormal{d}t}=\sum\limits_{m=1}^{N}\left(\gamma_{\Theta_m}^mp_{\Theta_m}(t)-\gamma_\Theta^m p_{\Theta}(t) \right) \;,
\label{eq:kin}
\end{equation}

\noindent where $p_{\Theta}(t)$ is the occupation probability of state $\Theta$, $\Theta_m$ is the network state obtained from $\Theta$ by flipping the state of $m$-th memristor, $\gamma_\Theta^m$ is the switching rate for $m$-th memristor in the configuration $\Theta$, and $\gamma_{\Theta_m}^m$ is defined similarly. The switching rate $\gamma_{\Theta}^m$ equals the switching probability (Eqs. (\ref{eq:gamma01}) or (\ref{eq:gamma10})) for $m$-th memristor in the state $\Theta$.

To demonstrate Eq.~(\ref{eq:kin}), consider two in-series connected identical memristors subjected to a voltage waveform $V_a(t)$. There are 4 possible network states that we denote as 00, 01, 10, and 11. In 00, both memristors are in the off-state, in 01, the first is in the off-, while the second is in the on-state, etc. Eq.~(\ref{eq:kin}) has the form

\begin{figure}
 \begin{center}
  (a)\includegraphics[width=1.4cm]{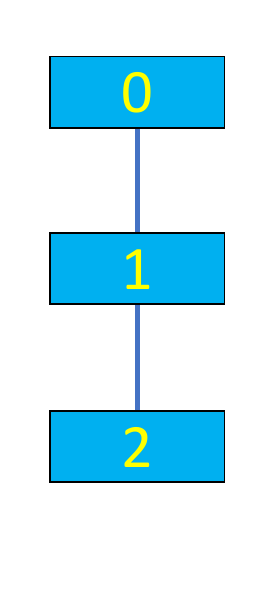} \hspace{1cm} (b)\includegraphics[width=4cm]{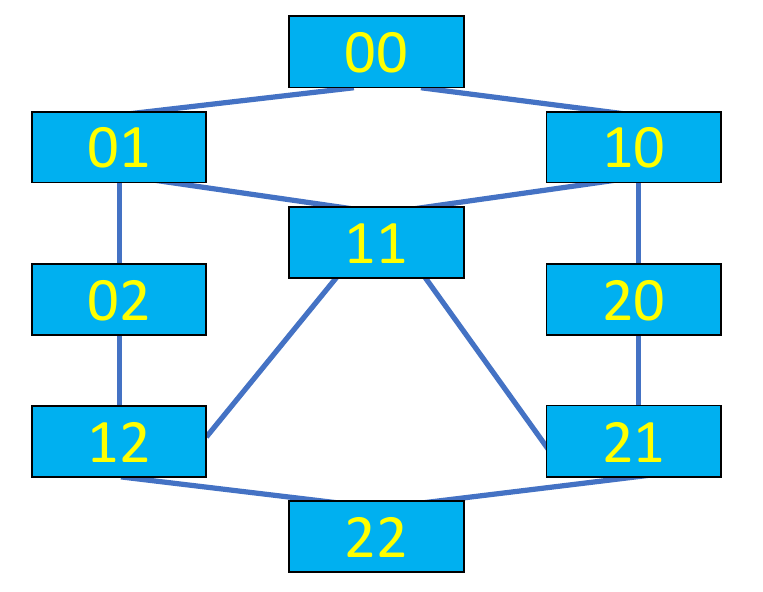}
  \fcaption{Transition scheme for (a) single three-state memristor, and (b) network of two three-state memristors.}
  \label{fig:1}
 \end{center}
 \vspace{0.5cm}
\end{figure}

\begin{eqnarray}
  \frac{\textnormal{d}p_{00}(t)}{\textnormal{d}t} &=& \gamma_{01}^1p_{01}+\gamma_{10}^2p_{10}-2\gamma_{00}^1p_{00}, \label{eq:twoM:1} \\
  \frac{\textnormal{d}p_{01}(t)}{\textnormal{d}t} &=& \gamma_{00}^1p_{00}+\gamma_{11}^2p_{11}-\gamma_{01}^2p_{01}-\gamma_{01}^1p_{01}, \;\;\;\;\label{eq:twoM:2} \\
  \frac{\textnormal{d}p_{10}(t)}{\textnormal{d}t} &=& \gamma_{00}^2p_{00}+\gamma_{11}^1p_{11}-\gamma_{10}^1p_{10}-\gamma_{10}^2p_{10}, \;\;\;\;\label{eq:twoM:3} \\
  \frac{\textnormal{d}p_{11}(t)}{\textnormal{d}t} &=& \gamma_{01}^2p_{01}+\gamma_{10}^1p_{10}-2\gamma_{11}^2p_{11}. \label{eq:twoM:4}
\end{eqnarray}

\noindent The similarity of memristors is taken into account by relations like $\gamma_{00}^1=\gamma_{00}^2$,
$\gamma_{01}^2=\gamma_{10}^1$, $p_{01}(t)=p_{10}(t)$, etc. Therefore, Eqs.~(\ref{eq:twoM:2}) and (\ref{eq:twoM:3}) are the same and the total number of equations that need to be solved reduces by one.
In our notation, $\gamma_{00}^1$ describes the switching rate from state 00 with the flipping of the 1-st memristor. The corresponding switching rate is given by Eq.~(\ref{eq:gamma01})
with $V=V_a(t)/2$, etc. Importantly, the computation of the switching rate involves the voltage across the switching memristor in the given configuration at the time moment $t$.\\

\rengSubsection{Multi-state memristors}

It is assumed that in a $K$-state memristor the switching between its boundary states ($R_{on}$ and $R_{off}$) occurs consecutively through $K-2$ intermediate resistance states.  The master equation (\ref{eq:kin}) preserves its form for multi-state memristor networks, but the network configuration space becomes more complex. Now the indices $i$, $j$, $k$, and so on, in the set $\Theta=(\dots kji)$ denoting the states of the first memristor, the second one, and so on, in the network can have more than two values. Generally, this leads to the exponential growth of the number of network states and, correspondingly, the number of independent equations for occupation probabilities $p_{\Theta}(t)$ when $N$, the number of memristors, increases. Luckily, the number of nonzero switching rates $\gamma$, corresponding to the nonzero terms in the right hand side of the master equation (\ref{eq:kin}) for a given network configuration $\Theta$, does not typically grow as fast.

In order to account for potential change in parameter values between resistance states, Eq.~(\ref{eq:gamma01}) and Eq.~(\ref{eq:gamma10}) are modified to
\begin{eqnarray}
\hspace{-0.5cm}  \gamma_{i\rightarrow j}(V)=\left\{ \begin{array}{cl}
\left( \tau_{ij} e^{-V/V_{ij}}\right)^{-1},& V>0,  j=i+1 \\
0 & \textnormal{otherwise}
\end{array}\right. \; , \label{eq:gammaij}\\
 \gamma_{j\rightarrow i}(V)=\left\{ \begin{array}{cl}
\left( \tau_{ji} e^{-|V|/V_{ji}}\right)^{-1},& V<0, j=i+1 \\
0 & \textnormal{otherwise}
\end{array}\right. \; . \label{eq:gammaji}
 \end{eqnarray}

\noindent with $\tau_{ij(ji)}$ and $V_{ij(ji)}$ being the constant values describing the resistance switching from $i(j)$-th to $j(i)$-th memristor state, and $i$ changes from 0 to $K-1$.

\begin{figure*}
 \centering
  (a)\includegraphics[width=1.5cm]{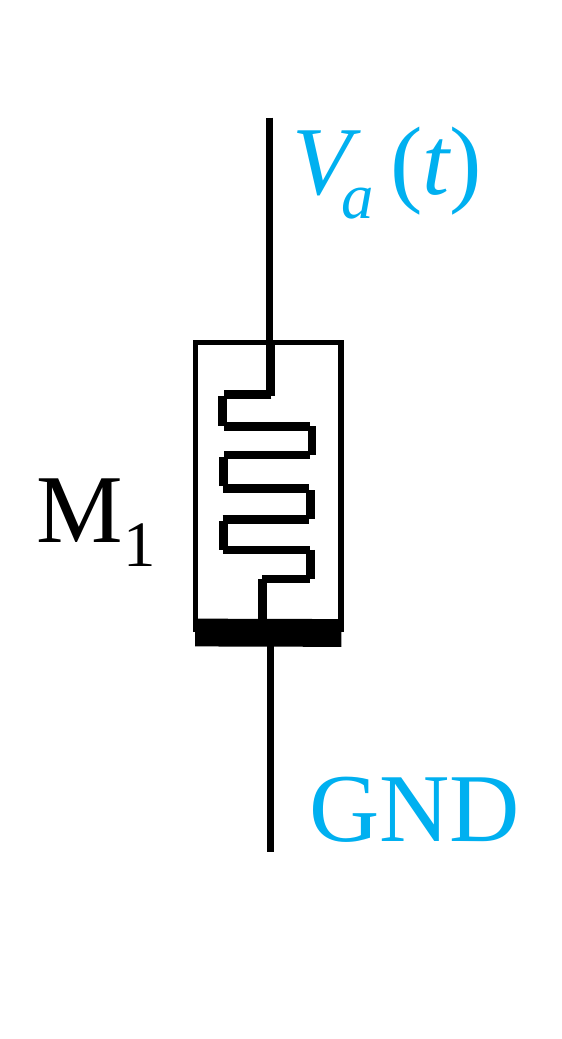} (b) \includegraphics[width=7cm]{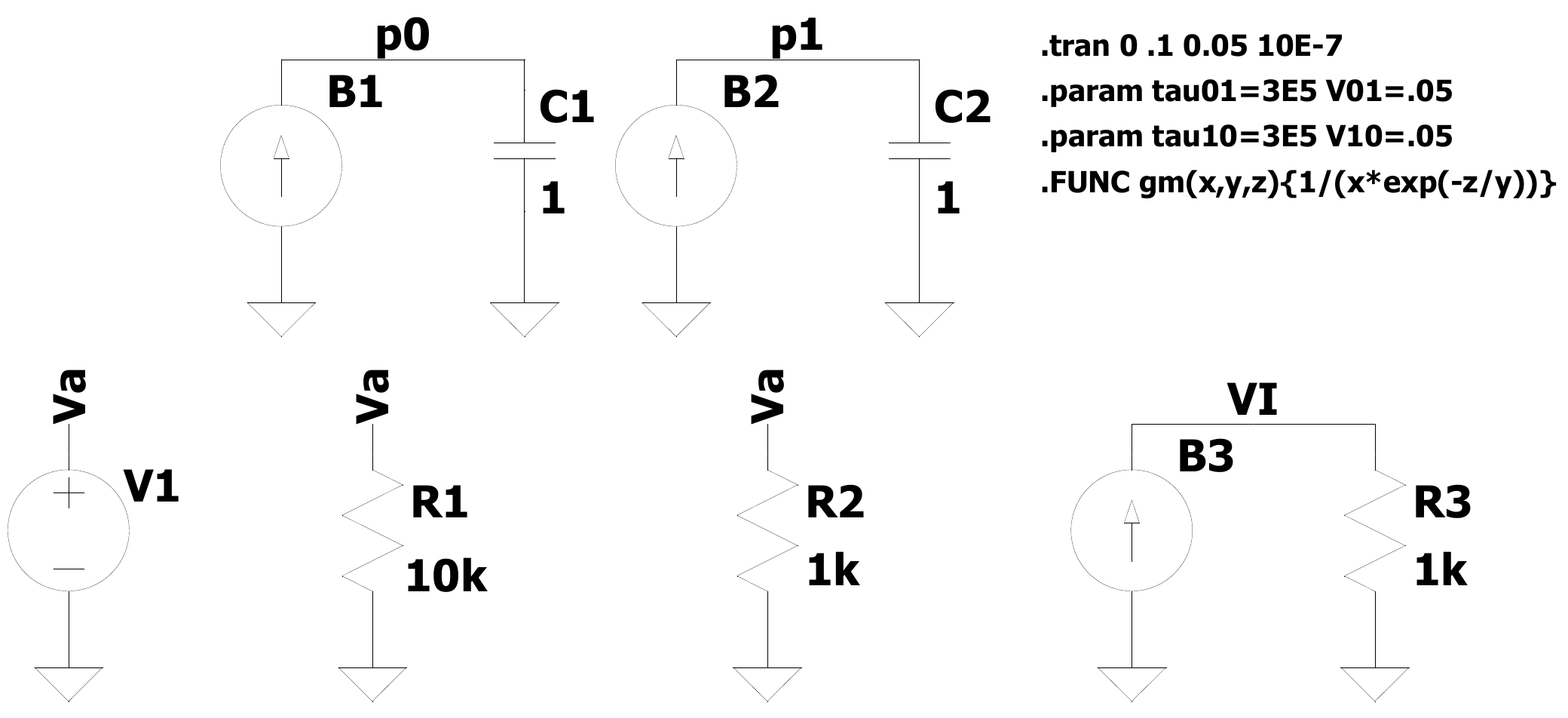} (c) \includegraphics[width=6cm]{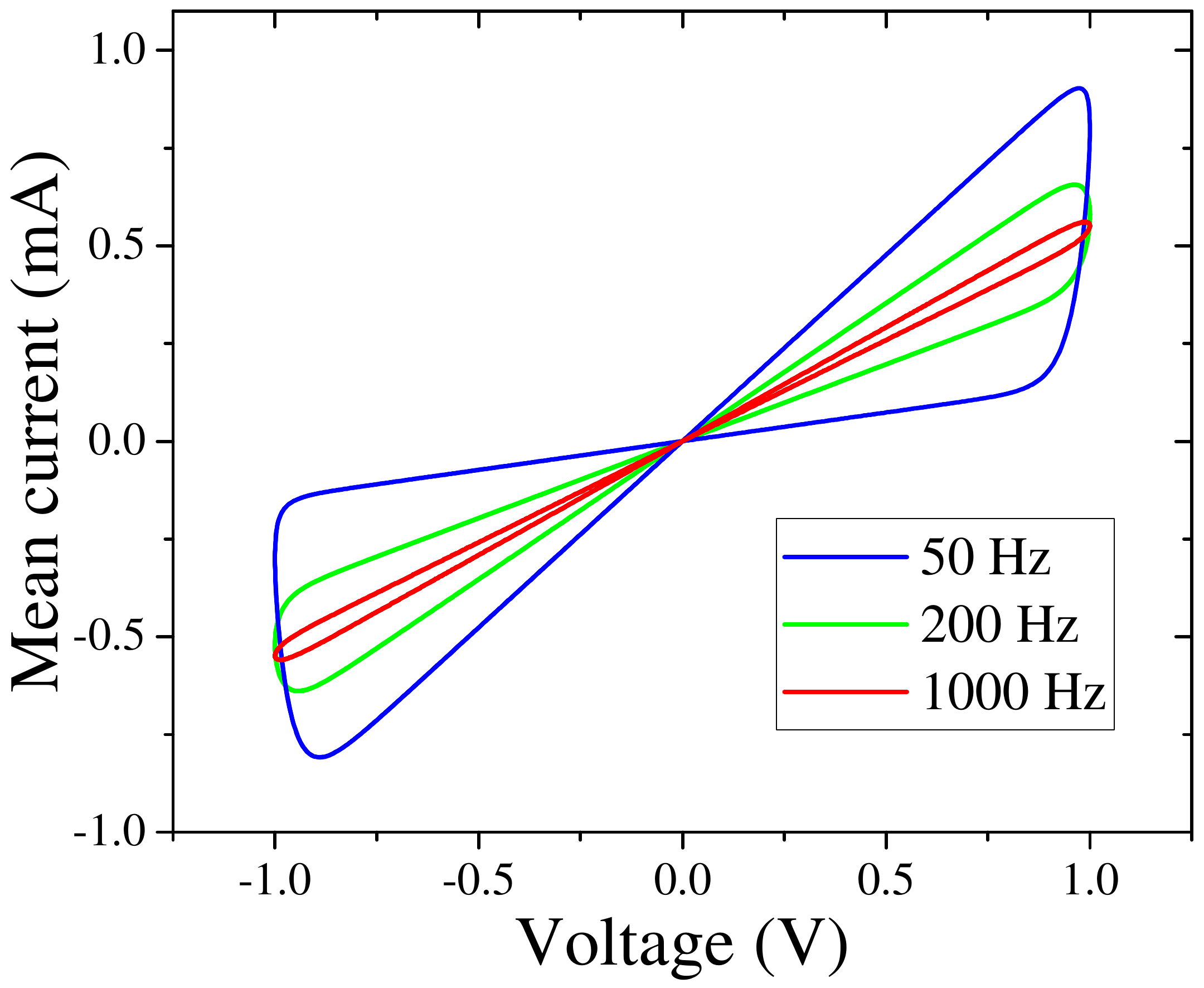}
  \fcaption{Ac-driven probabilistic binary memristor: (a) simulated circuit, (b) schematics of SPICE model, and (c) example of current-voltage curves found with SPICE simulations.  The listing of SPICE model is given in Table~\ref{tbl:1}.}
  \label{fig:2}
 %\end{center}
\end{figure*}
\begin{figure*}
 %\begin{center}
 \centering
  (a) \includegraphics[width=1.5cm]{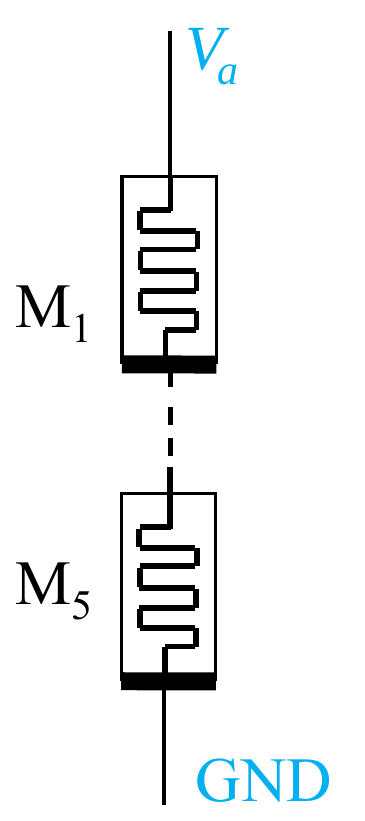} (b) \includegraphics[width=13cm]{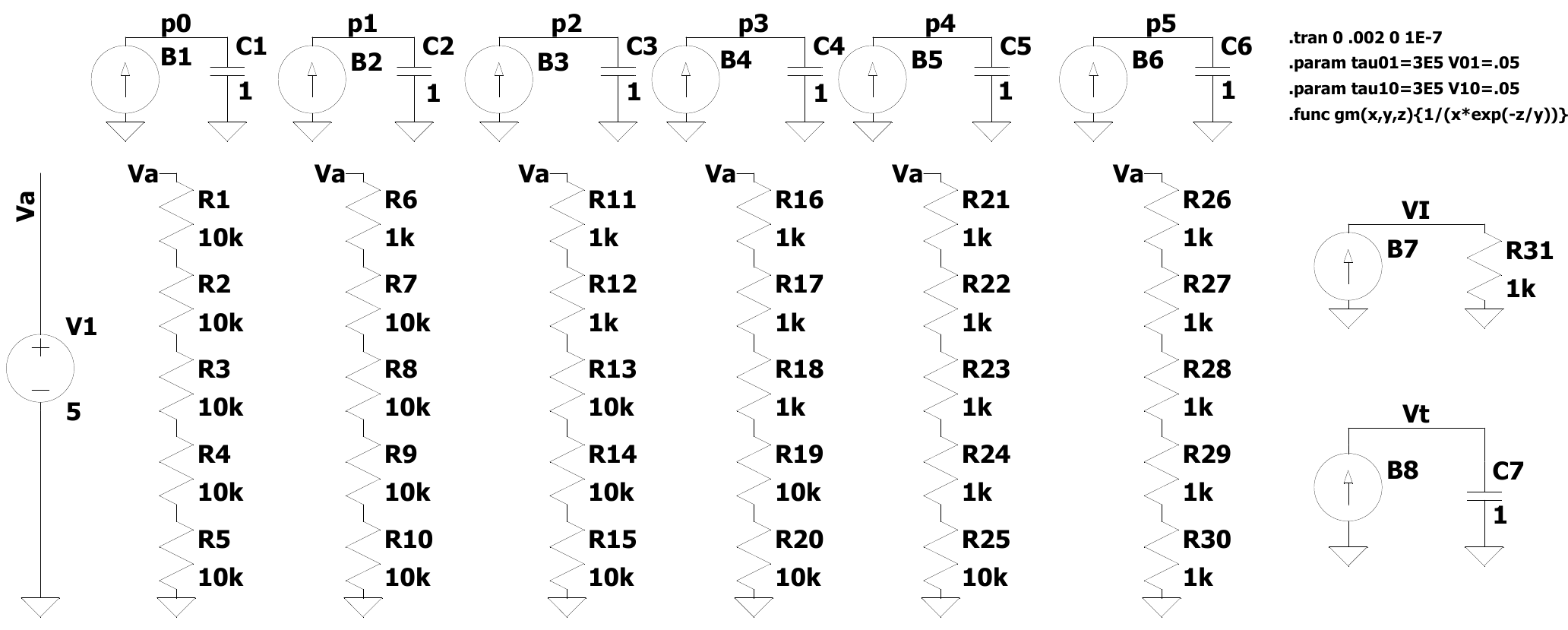}
  \fcaption{Dc-driven network of five probabilistic binary memristors: (a) simulated circuit, (b) schematics of SPICE model.}
  \label{fig:3}
 %\end{center}
\end{figure*}

It is convenient to represent the interdependencies between different occupation probabilities in the master equation using transition schemes. As an example, Figure (\ref{fig:1}) shows the transition schemes for a single three-state memristor (a) and two such memristors connected into network. An important feature of these schemes is the sequential change in the state of multi-state memristors that approximates the sequential growth of filaments in physical devices. Additionally, we emphasize that the transition schemes in general do not depend on the specific connections of memristors in the network. Such information is contained in the transition rates. In practice some of the transitions may be almost or entirely forbidden. For instance, when a positive voltage is applied to memristor described by Eqs. (\ref{eq:gamma01}) and (\ref{eq:gamma10}), the transition $1\rightarrow 0$ is forbidden  as it occurs at negative voltages. If one neglects low rate and/or forbidder transitions, we obtain the reduced transition scheme, which simplifies the solution of the master equation (\ref{eq:kin}) (see Ref.~\cite{dowling2020probabilistic} for some examples).

\rengSection{SPICE modeling approach}

Let $M$ be the number of non-equivalent equations for the occupation probabilites (like the set of Eqs.~(\ref{eq:twoM:1}), (\ref{eq:twoM:2}), and (\ref{eq:twoM:4})). The supremum of $M$ is $K^N$, where $K$ is the number of memristor states, and $N$ is the number of  memristors in the network. However, in practical cases $M$ can be much smaller than $K^N$. For instance, if there are $N$ binary ($K=2$) identical memristors connected in series, $M=K+1$ (see Ref.~\cite{dowling2020probabilistic}).

\begin{figure*}
  \begin{center}
  (a) \includegraphics[width=9cm]{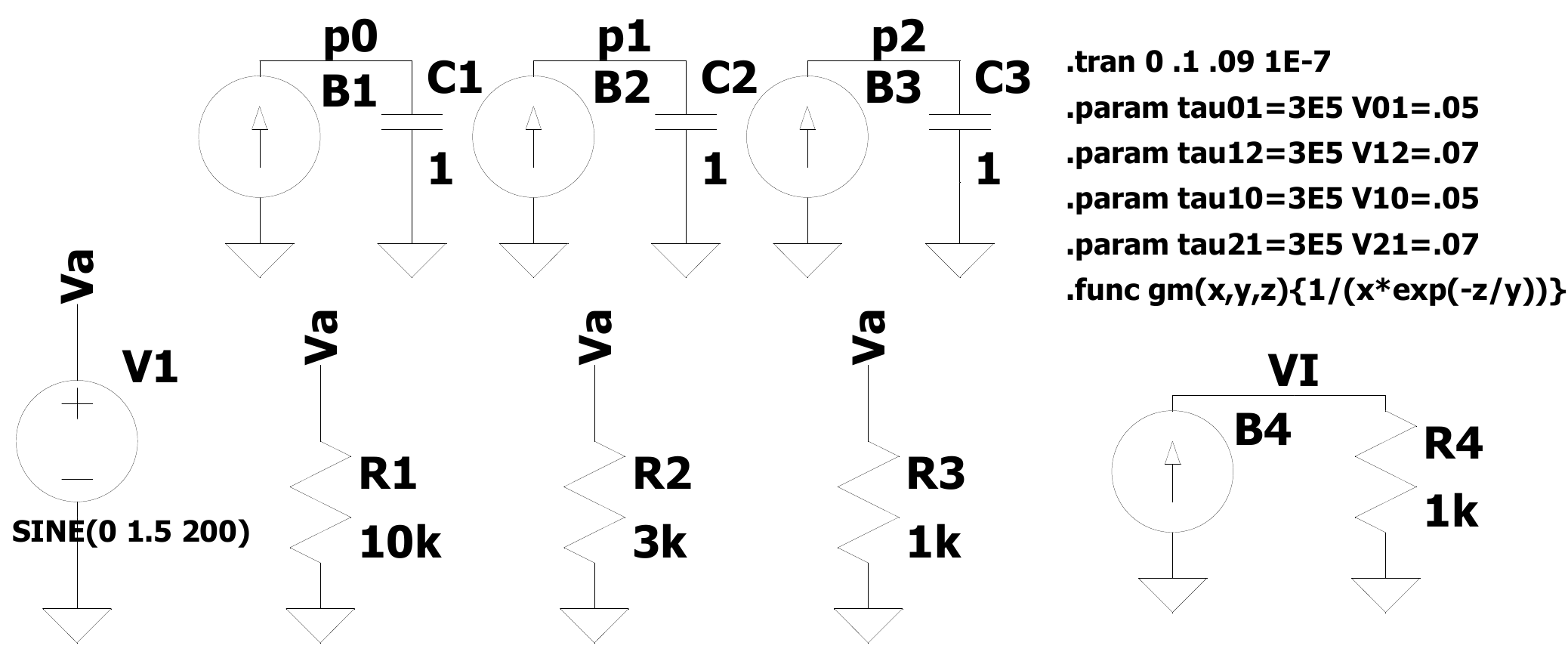}  (b) \includegraphics[width=6cm]{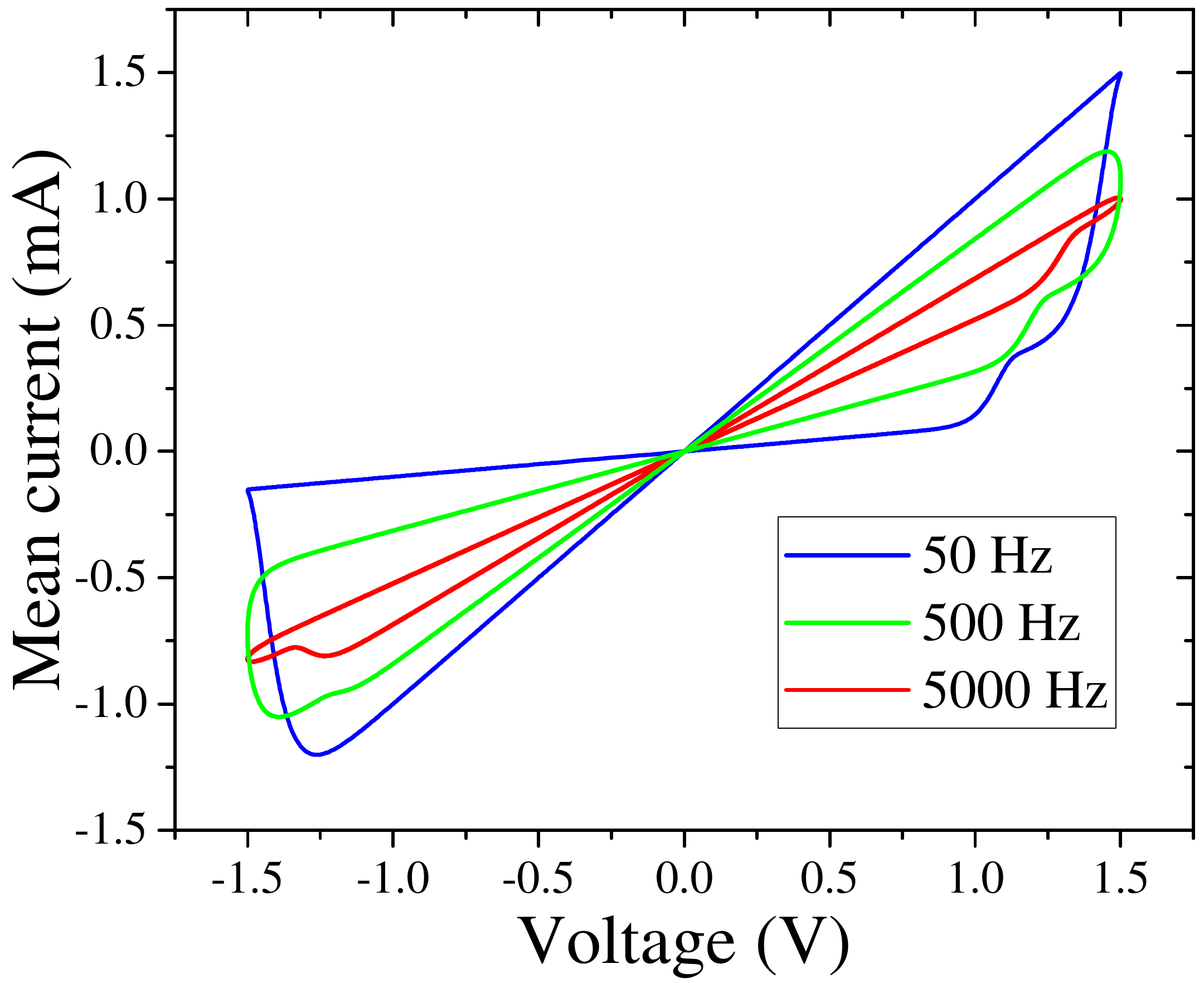}
  \fcaption{Ac-driven probabilistic three-state memristor: (a) schematics of SPICE model, and (c) example of current-voltage curves found with SPICE simulations.  The listing of SPICE model is given in Table~\ref{tbl:2}. The simulated circuit is the same as in Fig.~\ref{fig:2}(a) with the difference of different memristor type used.}\label{fig:5}
  \end{center}
\end{figure*}

\begin{figure*}
  \begin{center}
  (a) \includegraphics[width=1.5cm]{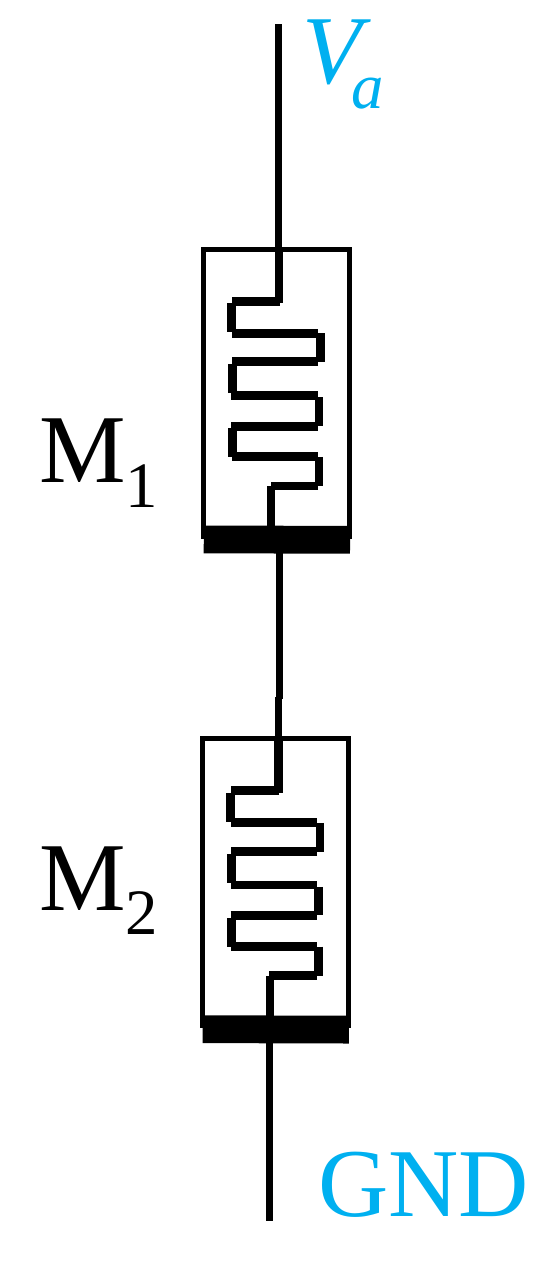} (b) \includegraphics[width=12cm]{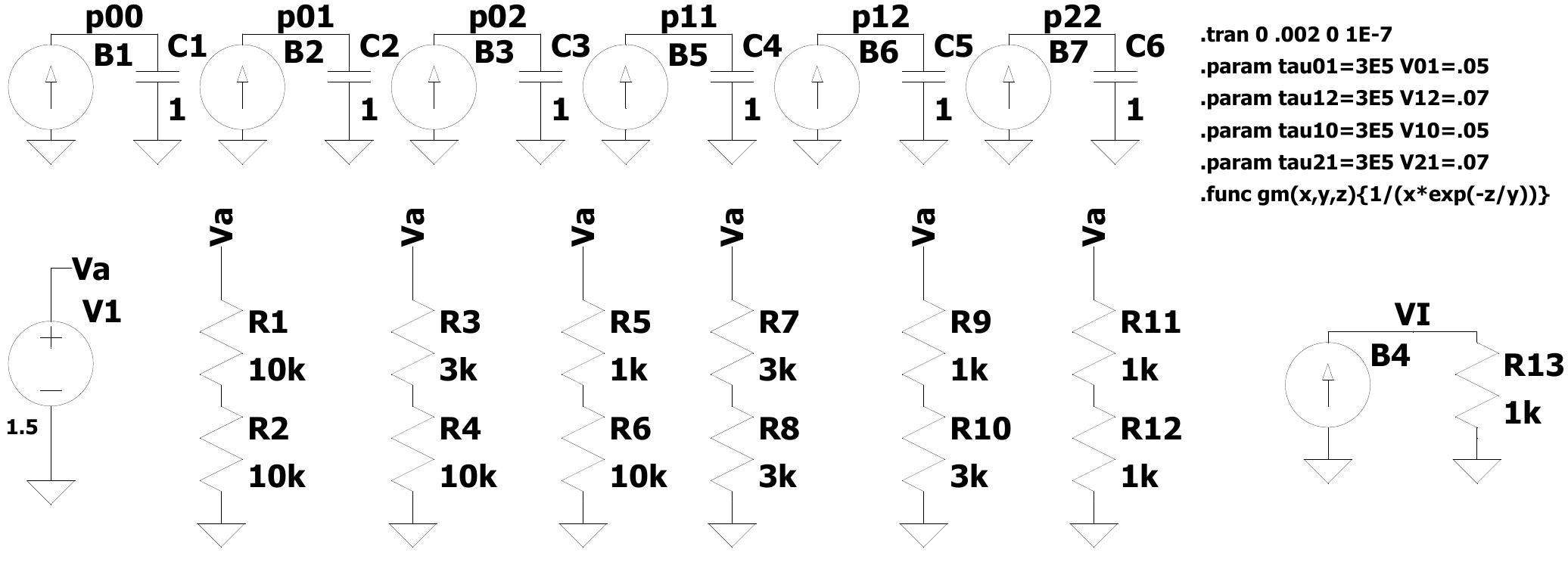} \\
  (c)  \includegraphics[width=6cm]{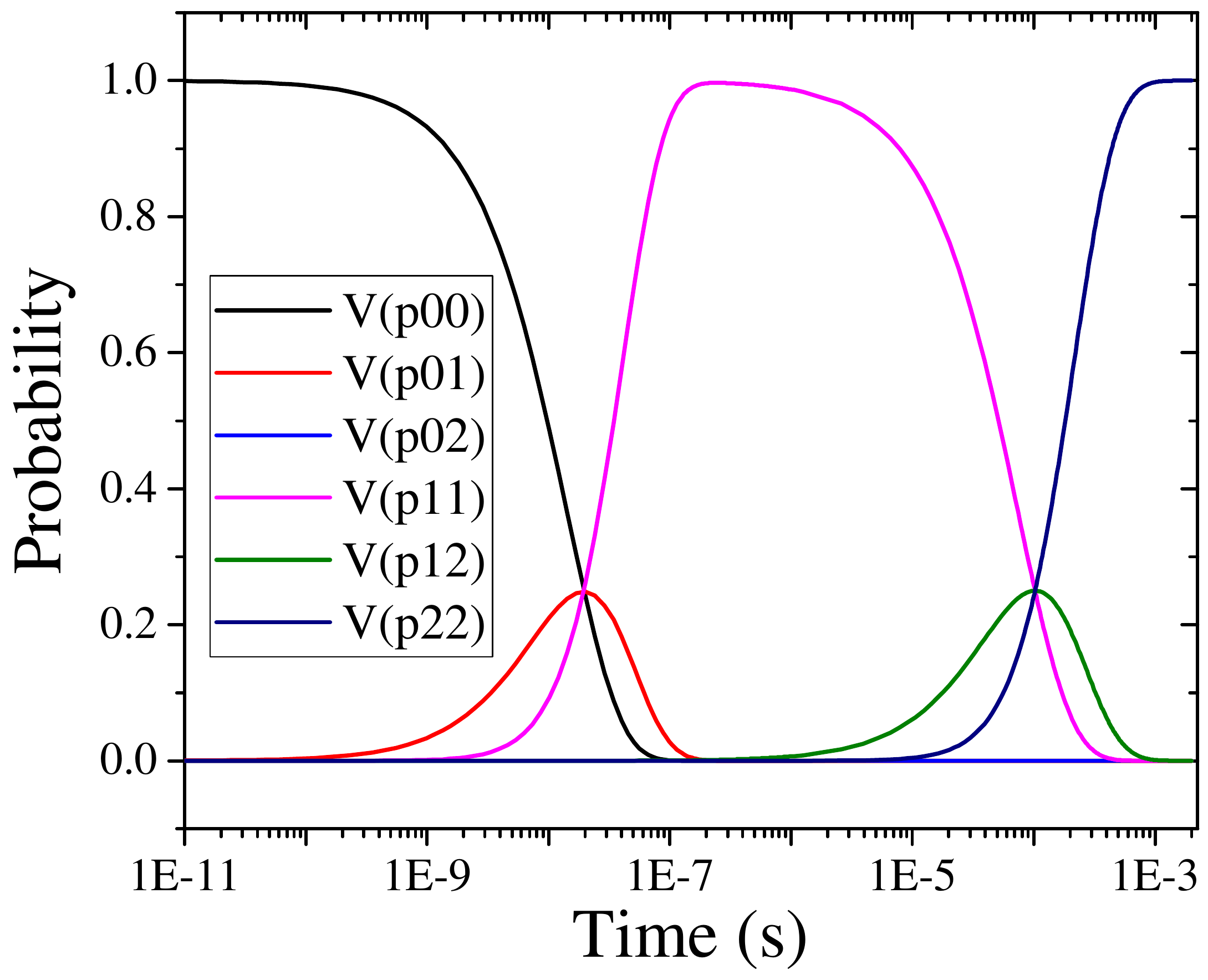} (d)  \includegraphics[width=6cm]{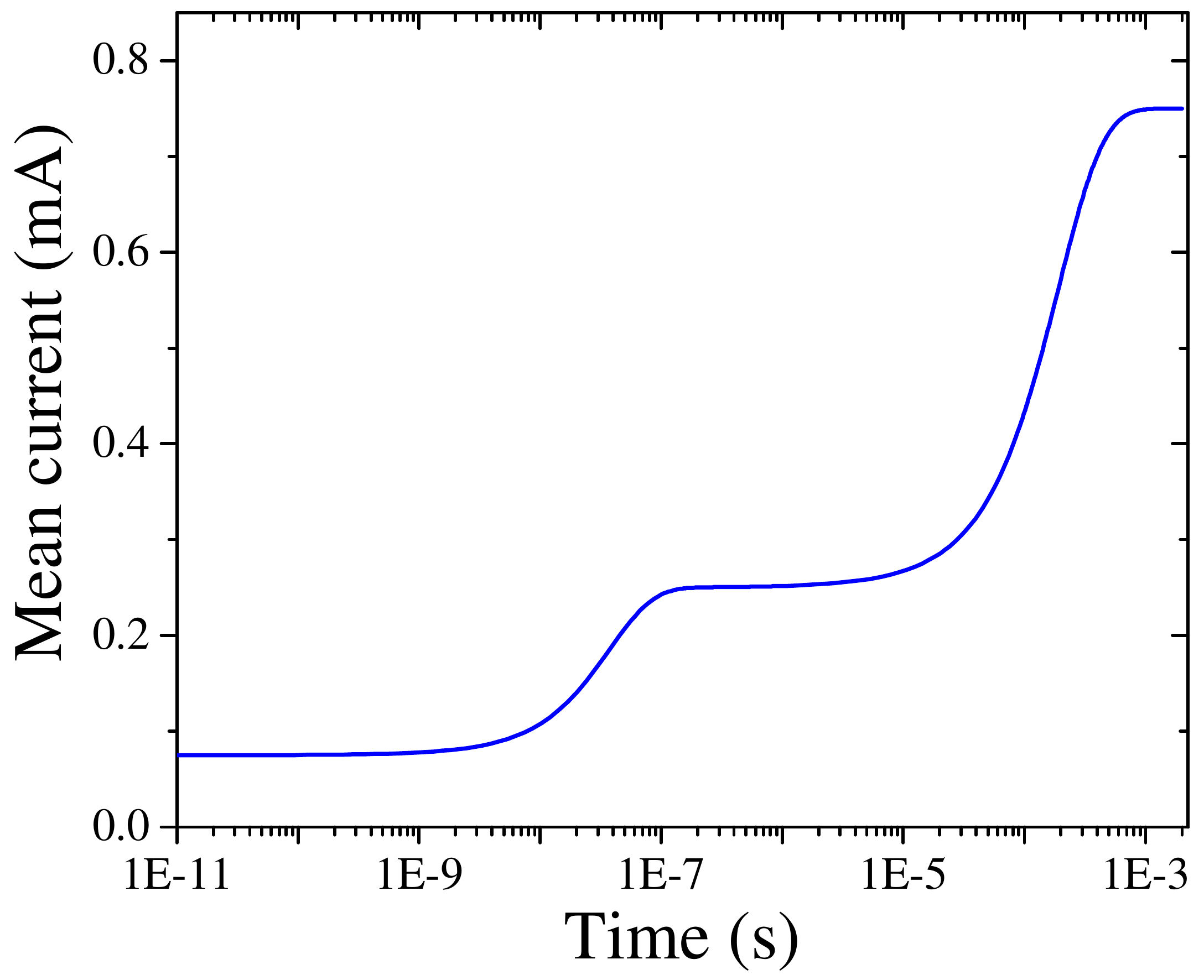}
  \fcaption{Dc-driven network of two three-state memristors: (a) simulated circuit, (b) schematics of SPICE model, (c) time-evolution of occupation probabilities, and (d) current as a function of time.}\label{fig:6}
  \end{center}
\end{figure*}

In the SPICE environment, we model each differential equation (such as Eq.~(\ref{eq:twoM:1})) by a 1 Farad capacitor charged by a voltage-controlled current source.  The occupation probabilities are represented by capacitor voltages. Each source current depends on the voltage across some of the capacitors which forms the right-hand side of the master equation. These circuits are shown in the top rows of SPICE models in  Figs.~\ref{fig:2}, \ref{fig:3}, \ref{fig:5} and \ref{fig:6}.

To account for the voltage-dependent switching rates (Eqs.~(\ref{eq:gamma01})-(\ref{eq:gamma10})),
$M$ copies of the network with memristors in non-equivalent combinations of states are utilized. These circuits (shown in the bottom row in Figs.~\ref{fig:2}, \ref{fig:3}, \ref{fig:5} and \ref{fig:6}) are connected to the input voltage. The voltages across memristors in these circuits are utilized to calculate the transition rates between the states.

To calculate the mean current, we use a voltage-controlled current source connected by a resistor to ground to provide a current path. For instance, in the case of in-series connected binary memristors, the current source output is defined by
\begin{equation}\label{eq:current}
\left< I \right> (t) \equiv   \sum\limits_{m=0}^{N} \begin{pmatrix}N\\m\end{pmatrix} I_m(t) p_m(t),
\end{equation}

\noindent where the number of states with the same number of memristors in the on-state is taken into account by the binomial coefficients $\begin{pmatrix}N\\m\end{pmatrix}$, and $I_m(t)$ is the current through the network with $m$ memristors in the on-state. The switching  time (or any other integral) can be evaluated numerically with a capacitor-voltage-controlled current source. Examples of such calculations can be found below.

\rengSection{Simulation examples}

\rengSubsection{AC-driven binary memristor}

In this simulation, a single binary memristor driven by an ac source is considered as seen in Fig.~\ref{fig:2}(a). Fig.~\ref{fig:2}(b) contains the schematic for the SPICE implementation and the corresponding SPICE code can be found in appendix A.1. The memristor has two possible states, $R_{on}$ and $R_{off}$, with resistance values of 1k and 10k Ohms respectively. We used the model parameter values $\tau_{01}=\tau_{10}=3\cdot 10^5$~s and $V_{01}=V_{10}=0.05$~V. The ac source,$V_a(t)$, has a peak voltage of 1V and is driven at various frequencies. The memristor is initialized in the off-state and will continue switching between the resistance states until the simulation has ended. The current is calculated using {\bf B4} and {\bf R4} components in Fig.~\ref{fig:2}(b). The current-voltage curves generated through SPICE simulation can be seen in Fig.~\ref{fig:2}(c) and they show the frequency behavior typical to deterministic memristive devices~\cite{chua76a,diventra09a}. We verified that
Fig.~\ref{fig:2}(b) SPICE model reproduces some previous results found through Monte Carlo simulations~\cite{dowling2020probabilistic}.  \\

\rengSubsection{DC-driven binary memristor network}

For this next simulation, we consider a network of binary memristors connected in-series as shown in Fig.~\ref{fig:3}(a). The network is composed of five memristors driven by a dc source with a voltage of 5V. Fig.~\ref{fig:3}(b) contains the schematic for the SPICE implementation. Each memristor is identical to one another, meaning the model parameters and the two states are equivalent from memristor to memristor. The memristors have two possible states, $R_{on}$ and $R_{off}$, with resistance values of 1k and 10k Ohms respectively. We used the model parameter values $\tau_{01}=\tau_{10}=3\cdot 10^5$~s and $V_{01}=V_{10}=0.05$~V.
\begin{figure}
 \begin{center}
   \includegraphics[width=0.8\columnwidth,keepaspectratio]{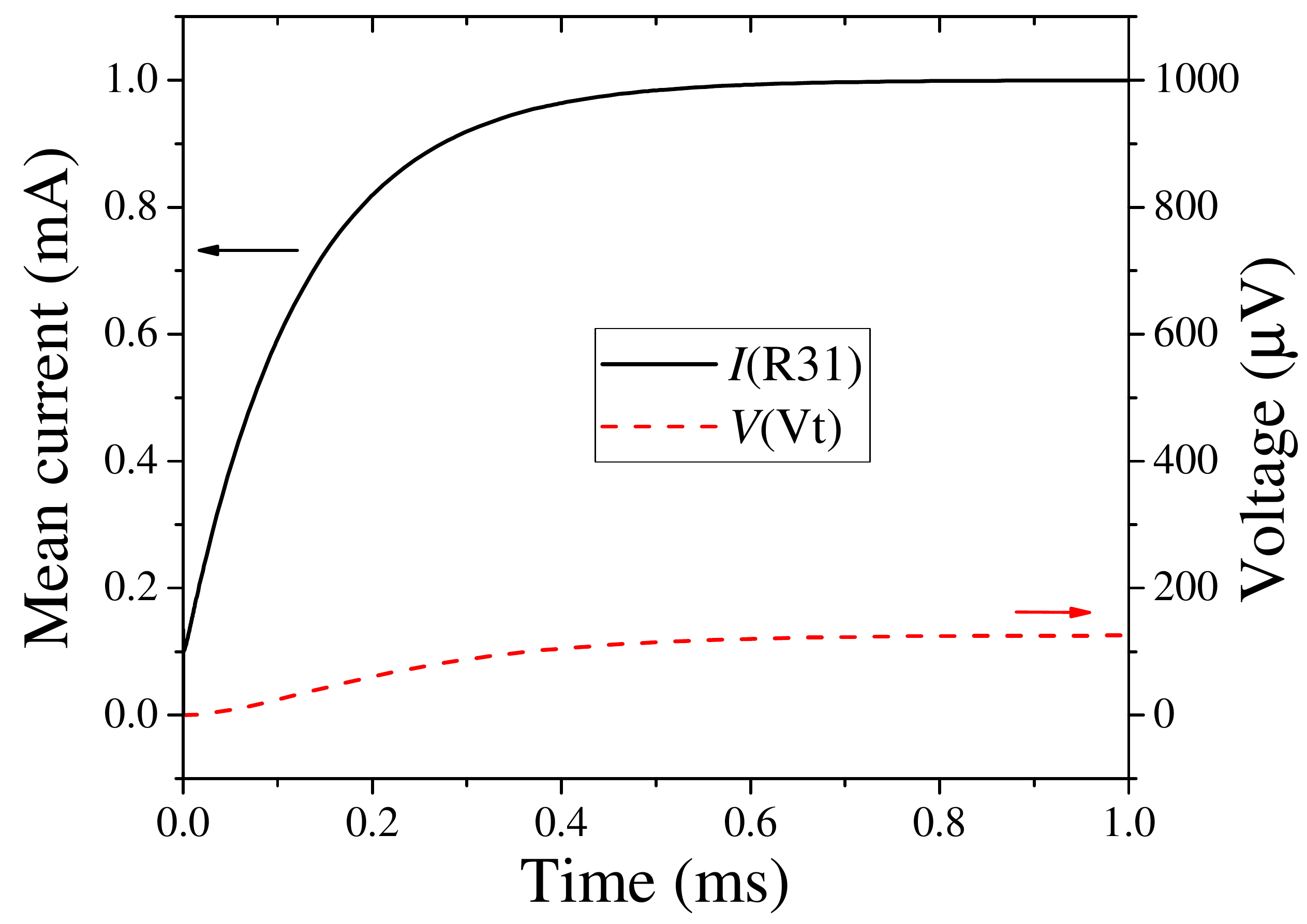}
   \fcaption{Current as a function of time (black solid line), and calculation of the network switching time (red dashed line) in the dc-driven network of five probabilistic binary memristors.}
   \label{fig:4}
  \end{center}
\end{figure}

\noindent Each memristor starts in the off-state and as time progresses each will switch to the on-state. When a memristor switches to the on-state, the drop in resistance causes an increase in the voltage across the off-state memristors increasing the probability of switching for the off-state memristor.

According to the analytical theory~\cite{dowling2020probabilistic}, the network mean switching time can be calculated as
\begin{equation}
\label{eq:App:sol13a}
  \left< T_N\right>=\sum_{j=0}^{N-1}\frac{1}{(N-j)\gamma_{j}}.
\end{equation}

\noindent For the parameters of simulations in Figs.~\ref{fig:3} and \ref{fig:4}, the above equation gives $\left< T_5\right>=126$~$\mu$s. Numerically, the same quantity can be evaluated using the following integral
\begin{equation}\label{eq:T2}
\int\limits_0^\infty t 5\gamma_{01111}^5p_{01111}(t) \textnormal{d}t.
\end{equation}

\noindent Technically, the integration is done by the components {\bf B8} and {\bf C7} in Fig.~\ref{fig:3}, so that the averaged switching time corresponds to the saturation limit of $V(\textnormal{Vt})$ curve in Fig.~\ref{fig:4}. We emphasize that the analytical and numerical (SPICE) values for $\left< T_5\right>$ are in full agreement.\\

\rengSubsection{Multi-state memristors}

The first multi-state simulation considered is a single tri-state memristor driven by an ac source. The ac source has a peak voltage of 1.5 V and is driven at various frequencies. Fig.~\ref{fig:5}(a) contains the schematic for the SPICE implementation and the corresponding SPICE code can be found in appendix A.2. The memristor now has three possible states, off-, intermediate, and on-state. To account for the added resistance state, a new copy of the memristor network is necessarily added to the SPICE implementation. These states have resistance values of 10k, 3k, and 1k Ohm respectively. The model parameters, $\tau_{ij}$ and $V_{ij}$, are as specified in the SPICE model schematics (Fig~\ref{fig:5}(a)). The memristor is initialized in the off-state and will continue switching between the resistance states until the simulation has ended. Fig.~\ref{fig:5}(b) shows the current-voltage curves generated by this SPICE simulation.

This next simulation is a network of two tri-state identical memristors driven by a 1.5V dc source shown in Fig.~\ref{fig:6}(a). The resistance states and model parameters are identical to the memristor used in the previous configuration. Fig.~\ref{fig:6}(b), the SPICE schematic used for this simulation is shown. The SPICE model is designed according to the transition scheme in Fig.~\ref{fig:1}(b). The memristors are initialized in the off-state and will switch to the intermediate state before switching to the on-state during the simulation. The evolution of resistance state probabilities for this network is shown in Fig.~\ref{fig:6}(c) and the mean current as a function of time for this SPICE simulation is shown in Fig.~\ref{fig:6}(d). The mean current increases in two steps because of the different time scales for the $0\rightarrow 1 $ and $1\rightarrow 2$
memristor switchings.

\rengSection{Summary}

In summary, the use of the master equation in probabilistic circuit modeling~\cite{dowling2020probabilistic} offers significant benefits compared to the routine Monte Carlo/stochastic simulations. Many circuit characteristics can be found on average in a single run and the master equation can be, in principle, solved analytically, with several analytical solutions already known~\cite{dowling2020probabilistic}. In this work, we have shown how to implement the master equation in SPICE. Our examples include simulations of binary and multi-state probabilistic memristors and their circuits subjected to ac- and dc-voltages. We expect that our approach will be useful to a broad range of researchers working in the area of emerging memory devices.

%\section*{Declaration of Competing Interest}

%The authors declare that they have no known competing financial interests or personal relationships that could have appeared to influence the work reported in this paper.




\bibliographystyle{IEEEtran}
\bibliography{memcapacitor}

\end{multicols}

\hspace{-0.5cm}

\rengAppendix{SPICE code examples}

\renewcommand\thetable{A.\arabic{table}}

\bigskip
\bigskip

\begin{table}
\noindent  B1 0 p0 I=-gm(tau01,V01,V(Va))*V(p0)*u(V(Va))+gm(tau10,V10,-V(Va))*V(p1)*u(-V(Va))\\
B2 0 p1 I=gm(tau01,V01,V(Va))*V(p0)**u(V(Va))-gm(tau10,V10,-V(Va))*V(p1)**u(-V(Va))\\
C1 p0 0 1 IC=1\\
C2 p1 0 1 IC=.0\\
R2 Va 0 1k\\
R1 Va 0 10k\\
R3 VI 0 1k\\
B3 0 VI I=I(R1)*V(p0)+I(R2)*V(p1)\\
V1 Va 0 SINE(0 1 200 0 0 0 0)\\
.FUNC gm(x,y,z){1/(x*exp(-z/y))}\\
.param tau01=3E5 V01=.05\\
.param tau10=3E5 V10=.05\\
.tran 0 .1 0.05 10E-7\\
.backanno\\
.end\\
\caption{\label{tbl:1} SPICE code for ac-driven probabilistic binary memristor.}
\end{table}

\bigskip
\bigskip

\begin{table}
\noindent B1 0 p0 I=(-gm(tau01,V01,V(Va))*V(p0))*u(V(Va))+(gm(tau10,V10,-V(Va))*V(p1))*u(-V(Va))\\
B2 0 p1 I=(gm(tau01,V01,V(Va))*V(p0)-gm(tau12,V12,V(Va))*V(p1))*u(V(Va))+(gm(tau21,V21,-V(Va))*V(p2)-gm(tau10,V10,-V(Va))*V(p1))*u(-V(Va))\\
B3 0 p2 I=(gm(tau12,V12,V(Va))*V(p1))*u(V(Va))+(-gm(tau21,V21,-V(Va))*V(p2))*u(-V(Va))\\
R1 Va 0 10k\\
R2 Va 0 3k\\
R3 Va 0 1k\\
R4 VI 0 1k\\
C1 p0 0 1 IC=1\\
C2 p1 0 1 IC=0\\
C3 p2 0 1 IC=0\\
B4 0 VI I=I(R1)*V(p0)+I(R2)*V(p1)+I(R3)*V(p2)\\
V1 Va 0 SINE(0 1.5 200)\\
.func gm(x,y,z){1/(x*exp(-z/y))}\\
.param tau01=3E5 V01=.05\\
.param tau12=3E5 V12=.07\\
.param tau10=3E5 V10=.05\\
.param tau21=3E5 V21=.07\\
.tran 0 .1 .09 1E-7\\
.backanno\\
.end\\
\caption{\label{tbl:2} SPICE code for ac-driven probabilistic three-state memristor.}
\end{table}

\end{document}